\documentclass[reqno]{amsart}
\newtheorem{thm}{Theorem}

       \theoremstyle{remark}
       \newtheorem{rem}[thm]{Remark}

\usepackage{amsfonts}
\usepackage[nosort]{cite}
\newcommand{\symp}[2]{#1\circ #2}
\newcommand{\algebra}[1]{\ensuremath{\mathfrak{#1}}}
\newcommand{\object}[2][\,]{\ensuremath{\mathrm{#2}#1}}
\newcommand{\Space}[3][]{\ensuremath{\mathbb{#2}^{#3}_{#1}{}}}
\newcommand{\such}{\,\mid\,}
\newcommand{\FSpace}[3][]{\ensuremath{#2_{#3}^{#1}{}}} 
\newcommand{\norm}[2][\relax]{\left\|#2\right\|\ifx#1\relax\else_{#1}\fi}
\newcommand{\modulus}[2][\relax]{\left| #2 \right|\ifx#1\relax\else_{#1}\fi}
\providecommand{\eqref}[1]{\textup{(\ref{#1})}}
\providecommand{\href}[2]{#2}
  \providecommand{\MR}[1]{\textbf{MR}~\href{http://www.ams.org/mathscinet-getitem?mr=#1}{\#~#1}}
\providecommand{\Zbl}[1]{\textbf{Zbl}~\href{http://www.emis.de:80/cgi-bin/zmen/ZMATH/en/zmathf.html?first=1&maxdocs=3&type=html&an=#1&format=complete}{\#~#1}}
\newcommand{\eprint}[2]{E-print: \href{#1}{\texttt{#2}}}
\def\ifundefined#1{\expandafter\ifx\csname#1\endcsname\relax}

\providecommand{\myhbar}{h}
\providecommand{\orbit}[1]{\mathcal{O}_{#1}}
\providecommand{\uir}[1]{\rho_{#1}}

\newcommand{\anti}{\mathcal{A}}

\newcommand{\ub}[3][]{\left\{\!#1\left[#2,#3\right]\!#1\right\}}

\newcommand{\ppnum}[3]{\gdef\rep@rtenum{#2}
  \gdef\rep@rteyear{#3}\gdef\wh@reappear{#1}}
\providecommand{\fl}{}

\providecommand{\keywords}[1]{}
\providecommand{\urladdr}[1]{}
\providecommand{\ams}[1]{\subjclass[2000]{#1}}
\providecommand{\rmi}{\mathrm{i}}

\ppnum{\href{http://arXiv.org/abs/math-ph/0306101}%
{arXiv:math-ph/0306101}
}{
}
{2003}

\begin{document}

\title{$p$-Mechanics and De~Donder--Weyl Theory}
\author[Vladimir V. Kisil]%
{\href{http://amsta.leeds.ac.uk/~kisilv/}{Vladimir
    V. Kisil}}
\thanks{On leave from the Odessa University.}
\email{kisilv@amsta.leeds.ac.uk}
\address{%
School of Mathematics,
University of Leeds,
Leeds LS2\,9JT,
UK}


\urladdr{\href{http://amsta.leeds.ac.uk/~kisilv/}%
{http://amsta.leeds.ac.uk/\~{}kisilv/}}

\begin{abstract}
  The orbit method of Kirillov is used to derive  
  the \(p\)-mechanical brackets~\cite{Kisil00a}. 
  They generate the quantum
  (Moyal) and classic (Poisson) brackets on respective orbits corresponding to
  representations of the Heisenberg group. 
  The extension of \(p\)-mechanics to field theory is made through the
  De~Donder--Weyl Hamiltonian formulation. The principal step is the
  substitution of the Heisenberg group with Galilean. 
\end{abstract}
\ams{81R05, 81R15, 81T70, 81T05, 81S10, 81S30, 22E27, 22E70, 43A65}

\maketitle


\section{Introduction}
\label{sec:introduction}

The purpose of this paper is to extended the \(p\)-mechanical
approach~\cite{Kisil96a,Kisil00a,Kisil02e,Brodlie03a,BrodlieKisil03a}
from particle mechanics to the field theory. We start in
Section~\ref{sec:preliminaries} from the Heisenberg group and its
representations derived through the orbit method of Kirillov. In
Section~\ref{sec:conv-algebra-hg} we define \(p\)-mechanical
observables as convolutions on the Heisenberg group \(\Space{H}{n}\)
and study their commutators. We modify the commutator of two
\(p\)-observables by the antiderivative to the central vector field in
the Heisenberg Lie algebra in Section~\ref{sec:p-mechanical-bracket},
this produces \(p\)-mechanical brackets and corresponding dynamic
equation. Then the \(p\)-mechanical construction is extended to the
De~Donder--Weyl Hamiltonian formulation of the field
theory~\cite{Kanatchikov95a,Kanatchikov98a,Kanatchikov98b,%
Kanatchikov98c,Kanatchikov98d,Kanatchikov99a,Kanatchikov00a,%
Kanatchikov01a,Kanatchikov01b} in
Section~\ref{sec:de-donder-weyl-1}. To this end we replace the
Heisenberg group by the Galilean group in
Section~\ref{sec:p-mech-appr}. Expanded presentation of
Section~\ref{sec:introduction-into-p} could
be found in~\cite{Kisil02e}. Development of material from
Section~\ref{sec:de-donder-weyl} will follow in subsequent papers.

\section{Elements of $p$-Mechanics}
\label{sec:introduction-into-p}

\subsection{The Heisenberg Group and Its Representations}
\label{sec:preliminaries}

Let \((s,x,y)\), where \(x\), \(y\in \Space{R}{n}\) and \(s\in\Space{R}{}\), be
an element of the Heisenberg group
\(\Space{H}{n}\)~\cite{Folland89,Howe80b}. The group law on
\(\Space{H}{n}\) is given as follows:
\begin{equation}
  \label{eq:H-n-group-law}
  (s,x,y)*(s',x',y')=(s+s'+\frac{1}{2} \omega(x,y;x',y'),x+x',y+y'), 
\end{equation} 
where the non-commutativity is made by \(\omega\)---the
\emph{symplectic form}~\cite[\S~37]{Arnold91} on \(\Space{R}{2n}\): 
\begin{equation}
  \label{eq:symplectic-form}
  \omega(x,y;x',y')=xy'-x'y.
\end{equation}
The Lie algebra \(\algebra{h}^n\) of \(\Space{H}{n}\) is spanned by
left-invariant vector fields 
\begin{equation}
  S={\partial_s}, \qquad
  X_j=\partial_{ x_j}-{y_j}/{2}{\partial s},  \qquad
 Y_j=\partial_{y_j}+{x_j}/{2}{\partial s}
  \label{eq:h-lie-algebra}
\end{equation}
on \(\Space{H}{n}\) with the Heisenberg \emph{commutator relations} 
\begin{equation}
  \label{eq:heisenberg-comm}
[X_i,Y_j]=\delta_{i,j}S
\end{equation}
and  all other commutators vanishing. 
The exponential map \(\exp:
\algebra{h}^n\rightarrow \Space{H}{n}\) respecting
the multiplication~\eqref{eq:H-n-group-law} and
Heisenberg commutators is
\begin{displaymath}
  \exp: sS+\sum_{k=1}^n (x_kX_k+y_kY_k) \mapsto (s,x_1,\ldots,x_n,y_1,\ldots,y_n).
\end{displaymath} 

As any group \(\Space{H}{n}\) acts on itself by the conjugation automorphisms
\(\object{A}(g) h= g^{-1}hg\), which fix the unit \(e\in
\Space{H}{n}\). The differential \(\object{Ad}: \algebra{h}^n\rightarrow
\algebra{h}^n\) of \(\object{A}\) at \(e\) is a linear map which could
be differentiated again to the representation \(\object{ad}\) of the Lie algebra
\(\algebra{h}^n\) by the
commutator: \(\object{ad}(A): B \mapsto [B,A]\). The dual space
\(\algebra{h}^*_n\) to the Lie algebra 
\(\algebra{h}^n\) is realised by the left invariant first order
differential forms on \(\Space{H}{n}\). By the duality
between \(\algebra{h}^n\) and \(\algebra{h}^*_n\) the map \(\object{Ad}\)
generates the \emph{co-adjoint
representation}~\cite[\S~15.1]{Kirillov76} \(\object{Ad}^*: \algebra{h}^*_n
\rightarrow \algebra{h}^*_n\):
\begin{equation}
  \label{eq:co-adjoint-rep}
  \object{ad}^*(s,x,y): (\myhbar ,q,p) \mapsto (\myhbar , q+\myhbar y,
  p-\myhbar x), \quad
  \textrm{where } (s,x,y)\in \Space{H}{n} 
\end{equation}
and \((\myhbar ,q,p)\in\algebra{h}^*_n\) in bi-orthonormal coordinates to
the exponential ones on \(\algebra{h}^n\).  There are two
types of orbits in~\eqref{eq:co-adjoint-rep} for \(\object{Ad}^*\)---Euclidean 
spaces \(\Space{R}{2n}\) and single points:
\begin{eqnarray}
  \label{eq:co-adjoint-orbits-inf}
  \orbit{\myhbar} & = & \{(\myhbar, q,p): \textrm{ for a fixed
  }\myhbar\neq 0 \textrm{ and  all } (q,p) \in  \Space{R}{2n}\}, \\
  \label{eq:co-adjoint-orbits-one}
  \orbit{(q,p)} & = & \{(0,q,p): \textrm{ for a fixed } (q,p)\in \Space{R}{2n}\}.
\end{eqnarray} The \emph{orbit method} of
Kirillov~\cite[\S~15]{Kirillov76} starts from the observation that the
above orbits parametrise irreducible unitary representations of
\(\Space{H}{n}\). All representations are
\emph{induced}~\cite[\S~13]{Kirillov76} by a character
\(\chi_\myhbar(s,0,0)=e^{2\pi \rmi \myhbar s}\) of the centre of
\(\Space{H}{n}\) generated by \((\myhbar,0,0)\in\algebra{h}^*_n\) and
shifts~\eqref{eq:co-adjoint-rep} from the \emph{left} on
orbits. Using~\cite[\S~13.2, Prob.~5]{Kirillov76} we get a neat
formula, which (unlike many other in literature) respects all
\emph{physical units}~\cite{Kisil02e}: 
\begin{equation}
  \textstyle
  \label{eq:stone-inf}
  \uir{\myhbar}(s,x,y): f_\myhbar (q,p) \mapsto 
  e^{ -2\pi\rmi( \myhbar s+qx+py)}
  f_\myhbar (q-\frac{\myhbar}{2} y, p+\frac{\myhbar}{2} x).
\end{equation}
The derived representation \(d\uir{\myhbar}\) of  the Lie algebra
\(\algebra{h}^n\) defined on the vector
fields~\eqref{eq:h-lie-algebra} is:
\begin{equation} 
  \textstyle 
  \fl
  d\uir{\myhbar}(S)=-2\pi\rmi  \myhbar I, \quad 
  d\uir{\myhbar}(X_j)=\myhbar \partial_{p_j}+ \frac{\rmi}{2}  q_j I ,\qquad
  d\uir{\myhbar}(Y_j)=-\myhbar \partial_{q_j}+ \frac{\rmi}{2}  p_j I
  \label{eq:der-repr-h-bar}
\end{equation}
Operators \(D^j_\myhbar\),  \(1\leq j\leq n\) representing vectors 
from the complexification of \(\algebra{h}^n\):
\begin{equation}
  \label{eq:Cauchy-Riemann}
  \textstyle \fl
  D^j_\myhbar =d\uir{\myhbar}(-X_j+ \rmi  Y_j)
  =\frac{\myhbar}{2}  (\partial_{p_j}+ \rmi \partial_{q_j})+2\pi(p_j+ \rmi  q_j) I
  ={\myhbar} \partial_{\bar{z}_j}+2\pi z_j I 
\end{equation} where \( z_j =p_j+ \rmi  q_j\) are used to give the
following classic result in terms of orbits:
\begin{thm}[Stone--von Neumann,
  cf. \textup{\cite[\S~18.4]{Kirillov76}, \cite[Chap.~1, \S~5]{Folland89}}] 
  \label{th:Stone-von-Neumann} 
  All unitary irreducible representations of \(\Space{H}{n}\) are
  parametrised up to equivalence by two classes of
  orbits~\eqref{eq:co-adjoint-orbits-inf}
  and~\eqref{eq:co-adjoint-orbits-one} of co-adjoint
  representation~\eqref{eq:co-adjoint-rep} in \(\algebra{h}^*_n\): 
  \begin{enumerate}
  \item The infinite dimensional representations by transformation
    \(\uir{\myhbar}\)~\eqref{eq:stone-inf} for \(\myhbar \neq 0\) in
    Fock~\textup{\cite{Folland89,Howe80b}} space
    \(\FSpace{F}{2}(\orbit{\myhbar})\subset\FSpace{L}{2}(\orbit{\myhbar})\)
    of null solutions to the operators \(D^j_\myhbar\)
    \eqref{eq:Cauchy-Riemann}:
    \begin{equation}
      \label{eq:Fock-type-space}
      \FSpace{F}{2}(\orbit{\myhbar})=\{f_{\myhbar}(p,q) \in
      \FSpace{L}{2}(\orbit{\myhbar}) \such D^j_\myhbar f_{\myhbar}=0,\
       1 \leq j \leq n\}.
    \end{equation}
  \item The one-dimensional representations as multiplication by
    a constant on \(\Space{C}{}=\FSpace{L}{2}(\orbit{(q,p)})\) which
    drops out from~\eqref{eq:stone-inf} for \(\myhbar =0\):
    \begin{equation}
      \label{eq:stone-one}
      \uir{(q,p)}(s,x,y): c \mapsto e^{-2\pi \rmi(qx+py)}c.
    \end{equation}
  \end{enumerate}
\end{thm} 
Note that \(f_\myhbar(p,q)\) is in
\(\FSpace{F}{2}(\orbit{\myhbar})\) if and only if the function
\(f_\myhbar(z)e^{-\modulus{z}^2/\myhbar}\), \(z=p+\rmi q\) is in the
classical Segal--Bargmann space~\cite{Folland89,Howe80b}, particularly
is analytical in \(z\). Furthermore the space
\(\FSpace{F}{2}(\orbit{\myhbar})\) is spanned by the Gaussian
\emph{vacuum vector} \(e^{-2\pi(q^2+p^2)/\myhbar}\) and all
\emph{coherent states}, which are ``shifts'' of the vacuum vector by
operators~\eqref{eq:stone-inf}.

Commutative representations~\eqref{eq:stone-one} correspond to the
case \(\myhbar=0\) in the formula~\eqref{eq:stone-inf}. They are always neglected,
however their union naturally (see the appearance of Poisson
brackets in~\eqref{eq:Poisson}) act as the classic \emph{phase space}:
\begin{equation}
  \label{eq:orbit-0}
  \orbit{0}=\bigcup_{(q,p)\in\Space{R}{2n}} \orbit{(q,p)}.
\end{equation}
Furthermore the structure of orbits of \(\algebra{h}_n^*\)
echoes in equation~\eqref{eq:p-equation} and its symplectic
invariance~\cite{Kisil02e}.  

\subsection{Convolution Algebra of $\Space{H}{n}$ and Commutator}
\label{sec:conv-algebra-hg}

Using  a left invariant measure \(dg\) on \(\Space{H}{n}\) the linear space
\(\FSpace{L}{1}(\Space{H}{n},dg)\)  can be upgraded 
to an algebra with the convolution multiplication:
\begin{equation}
  \fl
  (k_1 * k_2) (g) = \int_{\Space{H}{n}} k_1(g_1)\, k_2(g_1^{-1}g)\,dg_1 =
  \int_{\Space{H}{n}} k_1(gg_1^{-1})\, k_2(g_1)\,dg_1.
  \label{eq:de-convolution}
\end{equation}
Inner \emph{derivations} \(D_k\), \(k\in\FSpace{L}{1}(\Space{H}{n})\)
of \(\FSpace{L}{1}(\Space{H}{n})\) are given 
by the \emph{commutator} for \(f\in\FSpace{L}{1}(\Space{H}{n})\):
\begin{equation}
  \fl
  D_k: f \mapsto [k,f]=k*f-f*k
  =\int_{\Space{H}{n}} k(g_1)\left(
    f(g_1^{-1}g)-f(gg_1^{-1})\right)\,dg_1.
  \label{eq:commutator}
\end{equation}
A unitary representation \(\rho_\myhbar \) of \(\Space{H}{n}\) extends
 to \(\FSpace{L}{1}(\Space{H}{n} ,dg)\) by the formula:
\begin{eqnarray}
  \fl
\lefteqn{  [\rho_\myhbar (k)f](q,p)
 = \int_{\Space{H}{n}} k(g)\rho_\myhbar
  (g)f(q,p)\,dg \nonumber} \\
  &=&\int_{\Space{R}{2n}} \left(\int_{\Space{R}{}} k(s,x,y)e^{-2\pi\rmi \myhbar
      s}\,ds \right)
  e^{ -2\pi{\rmi}(qx+py)}      f (q-\myhbar y, p+\myhbar x) \,dx\,dy,
  \label{eq:rho-extended-to-L1}
\end{eqnarray}
thus \(\rho_\myhbar (k)\) for a fixed \(\myhbar \neq 0\) depends only
from \(\hat{k}_s(\myhbar,x,y)\)---the partial Fourier transform
\(s\rightarrow \myhbar\) of \(k(s,x,y)\). Then the representation of the composition of two
convolutions depends only from
\begin{eqnarray*}
  \fl
  (k'*k)\hat{_s}(\myhbar,x,y) 
  &=& 
  \int_{\Space{R}{2n}} e^{ \pi{\rmi \myhbar}  (xy'-yx')}\,
  \hat{k}'_s(\myhbar ,x',y')\,
  \hat{k}_s(\myhbar ,x-x',y-y')\,dx'dy'.
\end{eqnarray*}
The last expression for the full Fourier
transforms of \(k'\) and \(k\) turn to be the \emph{star product} known in {deformation
quantisation}, cf. \cite[(9)--(13)]{Zachos02a}.
Consequently the representation of  commutator~\eqref{eq:commutator}
depends only from~\cite{Kisil02e}:
\begin{eqnarray}
  \fl
    [k',k]\hat{_s}
 &=&   2 \rmi \! \int_{\Space{R}{2n}} \sin({\pi\myhbar}  (xy'-yx'))\,
 \hat{k}'_s(\myhbar ,x',y')
 \hat{k}_s(\myhbar ,x-x',y-y')\,dx'dy',
\label{eq:repres-commutator}
\end{eqnarray}
which turn to be  exactly the ``Moyal brackets''~\cite{Zachos02a} for the full Fourier
transforms of \(k'\) and \(k\). Also the expression~\eqref{eq:repres-commutator} 
vanishes  for \(\myhbar=0\) as can be expected from the commutativity of
representations~\eqref{eq:stone-one}.

\subsection{$p$-Mechanical Brackets on $\Space{H}{n}$}
\label{sec:p-mechanical-bracket}

A multiple \(\anti\) of a right inverse operator to the vector field
\(S\)~\eqref{eq:h-lie-algebra} on \(\Space{H}{n}\) is defined by:
\begin{equation}
  S\anti=4\pi^2 I, \qquad \textrm{ where }\quad
  \label{eq:def-anti}
  \anti e^{ 2\pi\rmi \myhbar s}=\left\{ 
    \begin{array}{ll}
      \frac{2\pi}{\rmi\myhbar} \strut e^{2\pi\rmi\myhbar s}, & \textrm{if } \myhbar\neq 0,\\
      4\pi^2 s, & \textrm{if } \myhbar=0.
    \end{array}
    \right. 
\end{equation} 
It can be extended by the linearity to
\(\FSpace{L}{1}(\Space{H}{n})\). We introduce~\cite{Kisil00a} a modified
convolution operation \(\star\) on \(\FSpace{L}{1}(\Space{H}{n})\) and
the associated modified commutator:
\begin{equation}
    \label{eq:star-and-brackets}
    k_1\star k_2= k_1*(\anti k_2), \qquad \ub{k_1}{k_2}=k_1\star
    k_2-k_2\star k_1.
\end{equation}
Then from~\eqref{eq:rho-extended-to-L1} one gets
\(\uir{\myhbar}(\anti k)=(i\myhbar)^{-1}\uir{\myhbar}(k)\) for
\(\myhbar\neq 0\). Consequently the modification
of~\eqref{eq:repres-commutator} for \(\myhbar\neq0\) is only slightly
different from the original one:
\begin{equation}
  \fl
    \ub{k'}{k}\!\hat{_s}
    =   \int_{\Space{R}{2n}}
    \frac{2\pi}{\myhbar}\sin(\pi\myhbar  (xy'-yx'))\,
    \hat{k}'_s(\myhbar ,x',y')\,
    \hat{k}_s(\myhbar ,x-x',y-y') \,dx'dy',
\label{eq:repres-ubracket}
\end{equation}
However the last expression for \(\myhbar=0\) is significantly distinct
from the vanishing~\eqref{eq:repres-commutator}. From the
natural assignment \(\frac{4\pi}{\myhbar}\sin(\pi\myhbar (xy'-yx'))=4\pi^2(xy'-yx')\) for
\(\myhbar=0\) we get the Poisson brackets for the Fourier
transforms of \(k'\) and \(k\) defined on \(\orbit{0}\)~\eqref{eq:orbit-0}:
\begin{equation}
  \label{eq:Poisson}
    \uir{(q,p)}\ub{k'}{k} = \frac{\partial \hat{k}'}{\partial q}
    \frac{\partial \hat{k}}{\partial p}
    -\frac{\partial \hat{k}'}{\partial p} \frac{\partial \hat{k}}{\partial q}.
\end{equation}
Furthermore the dynamical equation based on the modified
commutator~\eqref{eq:star-and-brackets} with a suitable Hamilton type
function \(H(s,x,y)\) for an observable \(f(s,x,y)\) on
\(\Space{H}{n}\) 
\begin{equation}
  \fl
  \label{eq:p-equation}
  \dot{f}=\ub{H}{f} \textrm{ is reduced } \left\{ 
    \begin{array}{l}
      \mbox{by \(\uir{\myhbar}\), \(\myhbar\neq0\) on
        \(\orbit{\myhbar}\)~\eqref{eq:co-adjoint-orbits-inf} to Moyal's
        eq. \cite[(8)]{Zachos02a};}\\
      \mbox{by \(\uir{(q,p)}\) on
        \(\displaystyle\orbit{0}\)~\eqref{eq:orbit-0} to Poisson's 
        eq. \cite[\S~39]{Arnold91}.}
    \end{array}
  \right.
\end{equation}
The same connections are true for the solutions of these three equations,
see~\cite{Kisil00a} for the harmonic oscillator
and~\cite{Brodlie03a,BrodlieKisil03a} for forced oscillator examples.

\section{De~Donder--Weyl Field Theory}
\label{sec:de-donder-weyl}

We extend \(p\)-mechanics to the De~Donder--Weyl field theory,
see~\cite{Kanatchikov95a,Kanatchikov98a,Kanatchikov98b,%
Kanatchikov98c,Kanatchikov98d,Kanatchikov99a,Kanatchikov00a,%
Kanatchikov01a,Kanatchikov01b}
for detailed exposition and further references. We will be limited
here to the preliminary discussion which extends the comment 5.2.(1)
from the earlier paper~\cite{Kisil02e}. Our notations will slightly
different from the used in the
papers~\cite{Kanatchikov95a,Kanatchikov98a,Kanatchikov98b,%
Kanatchikov98c,Kanatchikov98d,Kanatchikov99a,Kanatchikov00a,%
Kanatchikov01a,Kanatchikov01b}
to make it consistent with the used above and avoid clashes.

\subsection{Hamiltonian Form of Field Equation}
\label{sec:de-donder-weyl-1}

Let the underlying space-time have dimension and \(n+1\) parametrised
by coordinates \(u^\mu\), \(\mu=0,1,\ldots,n\) (with \(u^0\) parameter
traditionally associated with a time-like direction). Let a field be
described by \(m\) component tensor \(q^a\), \(a=1,\ldots,m\). For a
system defined by a Lagrangian density \(L(q^a, \partial_\mu q^a,
u^\mu)\) De~Donder--Weyl theory suggests new set of \emph{polymomenta}
\(p_a^\mu\) and \emph{DW Hamiltonian function}
\(H(q^a,p_a^\mu,u^\mu)\) defined as follows:
\begin{equation} \label{eq:polymomenta}
  p_a^\mu:=\frac{\partial L (q^a, \partial_\mu q^a,
u^\mu)}{\partial  (\partial_\mu y^a)}\quad
  \textrm{ and } \quad
  H(q^a,p_a^\mu,u^\mu)=p_a^\mu\,\partial_\mu q^a-L(q^a, \partial_\mu
  q^a, u^\mu). 
\end{equation}
Consequently the Euler--Lagrange field equations could be transformed
to the Hamil\-ton form:
\begin{equation}
  \label{eq:hamilton-field}
  \frac{\partial q^a}{\partial u^\mu}=\frac{\partial H}{\partial
    p_a^\mu}, \qquad
  \frac{\partial p_a^\mu}{\partial u^\mu}=-\frac{\partial H}{\partial
    q^a},
\end{equation} with the standard summation (over repeating indexes)
agreement. The main distinction from a particle mechanics is the
existence of \(n+1\) different polymomenta \(p_a^\mu\) associated to
each field variable \(q^a\). Correspondingly particle mechanics could
be considered as a particular case when \(n+1\) dimensional space-time
degenerates for \(n=0\) to ``time only''.

The next two natural
steps~\cite{Kanatchikov95a,Kanatchikov98a,Kanatchikov98b,%
Kanatchikov98c,Kanatchikov98d,Kanatchikov99a,Kanatchikov00a,%
Kanatchikov01a,Kanatchikov01b}
inspired by particle mechanics are:
\begin{enumerate}
\item Introduce an appropriate Poisson structure, such that the Hamilton
  equations~\eqref{eq:hamilton-field} will represent the Poisson
  brackets.
\item Quantise the above Poisson structure by some means,
  e.g. Dirac-Heisenberg-Shr\"odinger-Weyl technique or geometric
  quantisation.   
\end{enumerate}

We use here another path: first to construct a \(p\)-mechanical model
for equations~\eqref{eq:hamilton-field} and then deduce its quantum
and classical derivatives as was done for the particle mechanics
above. To simplify presentation we will start from the scalar field,
i.e. \(m=1\). Thus we drop index \(a\) in \(q^a\) and \(p_a^\mu\) and
simply write \(q\) and \(p^\mu\) instead.  

We also assume that underlying space-time is flat with a constant
metric tensor \(\eta^{\mu\nu}\). This metric define a related
\emph{Clifford algebra}~\cite{BraDelSom82,Cnops02a,DelSomSou92} with
generators \(e^\mu\) satisfying the relations
\begin{equation}
  \label{eq:clifford-def}
  e^\mu e^\nu + e^\nu e^\mu = \eta^{\mu\nu}.
\end{equation}
\begin{rem}
  For the Minkowski space-time (i.e. in the context of special
  relativity) a preferable choice is
  \emph{quaternions}~\cite{Sudbery79} with generators \(i\), \(j\),
  \(k\) instead the general Clifford algebra.
\end{rem}

Since \(q\) and \(p^\mu\) look like conjugated variables
\(p\)-mechanics suggests that they should generate a Lie algebra with
relations similar to~\eqref{eq:heisenberg-comm}. The first natural
assumption is the \(n+3(=1+(n+1)+1)\)-dimensional Lie algebra spanned
by \(X\), \(Y_\mu\), and \(S\) with the only non-trivial commutators
\([X,Y_\mu]=S\). However as follows from the Kirillov
theory~\cite{Kirillov62} any its unitary irreducible representation is
limited to a representation of \(\Space{H}{1}\) listed by the
Stone--von Neumann Theorem~\ref{th:Stone-von-Neumann}. Consequently
there is little chances that we could obtain the field
equations~\eqref{eq:hamilton-field} in this way.

\subsection{$p$-Mechanicanical Approach to the Field Theory}
\label{sec:p-mech-appr}

The next natural candidate is the Galilean group \(\Space{G}{n+1}\),
i.e. a nilpotent step 2 Lie group with
the \(2n+3(=1+(n+1)+(n+1))\)-dimensional
Lie algebra. It has a basis \(X\), \(Y_\mu\), and \(S_\mu\)
with \(n+1\)-dimensional centre spanned by \(S_\mu\). The only
non-trivial commutators  are
\begin{equation}
  \label{eq:galelean-comm}
  [X,Y_\mu]=S_\mu, \qquad \textrm{ where } \mu=0,1,\ldots,n.
\end{equation} Again the Kirillov theory~\cite{Kirillov62} assures
that any its \emph{complex valued} irreducible representation is a
representation of \(\Space{H}{1}\), but multidimensionality of the
centre offers an option~\cite{CnopsKisil97a,Kisil01d} to consider
\emph{Clifford valued} representations of \(\Space{G}{n+1}\). Thus we
proceed with this group.

\begin{rem}
  The appearance of Clifford algebra in connection with field theory
  and space-time geometry is natural. For example, the conformal
  invariance of space-time has profound consequences in
  astrophysics~\cite{Segal76} and, in their turn, conformal (M\"obius)
  transformations are most naturally represented by linear-fractional
  transformations in Clifford algebras~\cite{Cnops02a}. Some other
  links between nilpotent Lie groups and Clifford algebras are listed
  in~\cite{Kisil01d}.
\end{rem}

The Lie group \(\Space{G}{n+1}\) as manifold homeomorphic to
\(\Space{R}{2n+3}\) with coordinates \((s,x,y)\), where
\(x\in\Space{R}{}\) and \(s,y\in\Space{R}{n+1}\). The group
multiplication in this coordinates defined by,
cf.~\eqref{eq:H-n-group-law}:
\begin{eqnarray}
  \label{eq:Galilean-group-law}
  \lefteqn{(s,x,y)*(s',x',y')}\\
&=&(s_0+s_0'+\frac{1}{2}
  \omega(x,y_0;x',y_0'),\ldots, s_n+s_n'+\frac{1}{2}
  \omega(x,y_n;x',y_n'),x+x',y+y'). \nonumber 
\end{eqnarray} Observables are again defined as convolution operators
on \(\FSpace{L}{2}(\Space{G}{n+1})\). To define an appropriate
brackets of two observables \(k_1\) and \(k_2\) we will again modify
their commutator \([k_1,k_2]\) by antiderivative operators
\(\anti_0\), \(\anti_1\), \ldots, \(\anti_n\) which are multiples of
right inverse to the vector fields \(S_0\), \(S_1\), \ldots, \(S_n\),
cf.~\eqref{eq:def-anti}:
\begin{equation}
  S_\mu\anti_\mu=4\pi^2 I, \quad \textrm{ where }
  \label{eq:def-anti-Galilean}
  \anti_\mu e^{ 2\pi\rmi \myhbar s_\mu}=\left\{ 
    \begin{array}{ll}
      \frac{2\pi}{\rmi\myhbar} \strut e^{2\pi\rmi\myhbar s_\mu}, & \textrm{if } \myhbar\neq 0,\\
      4\pi^2 s_\mu, & \textrm{if } \myhbar=0,
    \end{array}
    \right. \textrm{ and } \mu=0,1,\ldots,n.
\end{equation} 
The definition of the brackets follows the ideas
of~\cite[\S~3.3]{CnopsKisil97a}: to each vector field \(S_\mu\)
should be associated a generator \(e^\mu\) of Clifford
algebra~\eqref{eq:clifford-def}.  
Thus our brackets are as follows, cf.~\eqref{eq:star-and-brackets}:
\begin{equation}
  \label{eq:field-brackets}
  \ub{B_1}{B_2}=B_1* \anti B_2 -
  B_2*\anti B_1, \qquad \textrm{ where }  \anti=e^\mu\anti_\mu.
\end{equation}
These brackets will be used in the right-hand side of the \(p\)-dynamic
equation. Its left-hand side should contain a replacement for the time
derivative. As was already mentioned
in~\cite{Kanatchikov95a,Kanatchikov98a,Kanatchikov98b,%
Kanatchikov98c,Kanatchikov98d,Kanatchikov99a,Kanatchikov00a,%
Kanatchikov01a,Kanatchikov01b}
the space-time play a r\^ole of multidimensional time in the
De~Donder--Weyl construction. Thus we replace time derivative by the
symmetric pairing \(\symp{D}{}\) with the \emph{Dirac
  operator}~\cite{BraDelSom82,Cnops02a,DelSomSou92} \(D=e^\mu
\partial_\mu\) as follows:
\begin{equation}
  \label{eq:dirac}
  \symp{D}{f}=-\frac{1}{2}\left(e^\mu \frac{\partial f}{\partial
      u^\mu}  + \frac{\partial f}{\partial u^\mu} e^\mu \right),
  \qquad \textrm{ where } D=e^\mu \partial_\mu. 
\end{equation}
Finally the \(p\)-mechanical dynamic equation, cf.~\eqref{eq:p-equation}:
\begin{equation}
  \label{eq:p-mech-field-eq}
  \symp{D}{f}=\ub{H}{f},
\end{equation}
is defined through the brackets~\eqref{eq:field-brackets} and the
Dirac operator~\eqref{eq:dirac}.

To ``verify'' the equation~\eqref{eq:p-mech-field-eq} we will find its
classical representation and compare it
with~\eqref{eq:hamilton-field}. Similarly to calculations in
section~\ref{sec:conv-algebra-hg} we find, cf.~\eqref{eq:Poisson}:
\begin{equation}
  \label{eq:field-Poisson}
    \uir{(q,p^\mu)}\ub{k'}{k} = \frac{\partial \hat{k}'} {\partial q}e^\mu
    \frac{\partial \hat{k}}{\partial p^\mu}
    -\frac{\partial \hat{k}'}{\partial p^\mu} e^\mu \frac{\partial \hat{k}}{\partial q}.
\end{equation}
Consequently the dynamic of field observable \(q\) from the
equation~\eqref{eq:p-mech-field-eq} with a
scalar-valued Hamiltonian \(H\) is given by:
\begin{equation}
  \label{eq:field-q-eq}
  \symp{D}{q}= \left(\frac{\partial H} {\partial q}e^\mu
    \frac{\partial}{\partial p^\mu}
    -\frac{\partial H}{\partial p^\mu} e^\mu \frac{\partial
      }{\partial q}\right) q \qquad \Longleftrightarrow  \qquad 
    \frac{\partial q}{\partial u^\mu}e^\mu = \frac{\partial H}{\partial p^\mu} e^\mu,
\end{equation}
i.e. after separation of components with different generators
\(e^\mu\) we get first \(n+1\) equations from~\eqref{eq:hamilton-field}.

To get the last equation for polymomenta~\eqref{eq:hamilton-field} we
again use the Clifford algebra generators to construct the
\emph{combined polymomenta} \(p=e_\nu p^\nu\). For them:
\begin{eqnarray*}
  \symp{D}{p} &=& -\frac{1}{2}\left(e^\mu \frac{\partial e_\nu p^\nu}{\partial
      u^\mu}  + \frac{\partial e_\nu p^\nu}{\partial u^\mu} e^\mu
  \right) = -\frac{\partial p^\mu}{\partial u^\mu} e^\mu e_\mu, \\
  \ub{H}{f} &=& \frac{\partial H} {\partial q}e^\mu
    \frac{\partial e_\nu p^\nu}{\partial p^\mu}
    -\frac{\partial H}{\partial p^\mu} e^\mu \frac{\partial
      e_\nu p^\nu}{\partial q} = \frac{\partial H} {\partial q}e^\mu e_\mu.
\end{eqnarray*}
Thus the equation~\eqref{eq:p-mech-field-eq} for the combined polymomenta
\(p=e_\nu p^\nu\) becomes: 
\begin{equation}
  \label{eq:field-p-eq}
    \frac{\partial p^\mu}{\partial u^\mu} e^\mu e_\mu =
    -\frac{\partial H}{\partial q} e^\mu e_\mu, 
\end{equation}
i.e. coincides with the last equation in~\eqref{eq:hamilton-field} up
to a constant factor \(e^\mu
e_\mu\). 

Consequently images of the equation~\eqref{eq:p-mech-field-eq} under
the infinite dimensional representation of the group
\(\Space{G}{n+1}\) could stand for quantisations of its classical
images in~\eqref{eq:hamilton-field}, \eqref{eq:field-Poisson}. 
A further study of quantum images of the
equation~\eqref{eq:p-mech-field-eq} as well as extension to vector or
spinor fields should follow in subsequent papers.

\begin{rem}
  To consider vector or spinor fields with components \(q_a\),
  \(a=1,\ldots,m\) it worths to introduce another Clifford algebra
  with generators \(c^a\) and consider a composite field
  \(q=c^aq_a\). There are different ways to link Clifford and Grassmann
  algebras, see e.g.~\cite{Berezin86,Hestenes99}. Through such a link
  the Clifford algebra with generators \(e_\mu\) corresponds to
  horizontal differential forms in the sense
  of~\cite{Kanatchikov95a,Kanatchikov98a,Kanatchikov98b,%
Kanatchikov98c,Kanatchikov98d,Kanatchikov99a,Kanatchikov00a,%
Kanatchikov01a,Kanatchikov01b}
  and the Clifford algebra generated by \(c_a\)---to the vertical.
\end{rem}

\subsection*{Acknowledgement} 
I am very grateful to
Prof.~I.V.~Kanatchikov for introducing me to the De~Donder--Weyl
theory and many useful discussions and comments.


\newcommand{\noopsort}[1]{} \newcommand{\printfirst}[2]{#1}
  \newcommand{\singleletter}[1]{#1} \newcommand{\switchargs}[2]{#2#1}
  \newcommand{\irm}{\textup{I}} \newcommand{\iirm}{\textup{II}}
  \newcommand{\vrm}{\textup{V}}
  \providecommand{\cprime}{'}\providecommand{\arXiv}[1]{\eprint{http://arXiv.o%
rg/abs/#1}{arXiv:#1}}
\providecommand{\bysame}{\leavevmode\hbox to3em{\hrulefill}\thinspace}
\providecommand{\MR}{\relax\ifhmode\unskip\space\fi MR }
\providecommand{\MRhref}[2]{%
  \href{http://www.ams.org/mathscinet-getitem?mr=#1}{#2}
}
\providecommand{\href}[2]{#2}

\end{document}